\documentclass[twocolumn,amsmath,amssymb,floatfix]{revtex4}

\usepackage{graphicx}
\usepackage{gensymb} 
\usepackage{epstopdf}

\begin{document}

\newcommand{\Jc}{J_\text{c}}
\newcommand{\Ha}{H_\text{a}}
\newcommand{\Hc}{H_\text{c}}
\newcommand{\Hu}{H_\uparrow}
\newcommand{\Hd}{H_\downarrow}
\newcommand{\Jd}{J_\downarrow}
\newcommand{\xp}{x_\text{p}}
\newcommand{\Hl}{H_\text{l}}
\newcommand{\Bo}{B_\text{0}}
\newcommand{\Jco}{J_\text{c0}}

\renewcommand{\thefootnote}{}

\title{Diamond Magnetometry of Meissner Currents in a Superconducting Film}
\author{Nir Alfasi*\footnotetext{*These authors contributed equally to this work}, Sergei Masis*, Oleg Shtempluck, Valleri Kochetok, Eyal Buks}

\affiliation{Andrew and Erna Viterbi Department of Electrical Engineering, Technion, Haifa 32000 Israel}

\date{\today}

\begin{abstract}

We study magnetic field penetration into a thin film made of a superconducting niobium. Imaging of magnetic field is performed by optically detecting magnetic resonances of negatively charged nitrogen-vacancy defects inside a single crystal diamond, which is attached to the niobium film under study. The experimental results are compared with theoretical predictions based on the critical state model, and good agreement is obtained.

\end{abstract}

\maketitle

Magnetic field imaging is widely employed in the study of superconductors \cite{Bending_449}. A variety of techniques, including magnetic force
microscopy \cite{Nakano_100510,Mamin_301}, Hall sensing \cite{Chang_1974,Lyard_057001,Lyard_180504,Okazaki_064520,Zeldov_373,Beek_174517,Abulafia_2891,Oral_1202}, magneto-optical imaging \cite{Beek_174517,Dorosinskii_149,Johansen_16264,Jooss_651,Soibel_282} and scanning superconducting quantum interference device magnetometry \cite{Kirtley_117,Hasselbach_4432,kuit_134504,Embon_7598,Vasyukov_639} have been used to perform spatially resolved measurements of magnetic properties of superconductors \cite{Lyard_057001,Lyard_180504,Okazaki_064520,Nakano_100510,Luan_067001,Shapoval_214517,Zeldov_373,Beek_174517}.

Here we employ diamond-based vectorial magnetometry for imaging the penetration of magnetic field into a type II superconductor. A cryogenic magnetometer that allows optical detection of magnetic resonance (ODMR) is employed for imaging the penetration as a function of externally applied magnetic field. The comparison between the experimental findings and theoretical predictions based on the Bean critical state model \cite{Bean_31,Campbell_199,Brandt_735,Zeldov_9802,Fietz_A335} yields a good agreement.

\begin{figure}[b]
	\centering
	\includegraphics[width=0.98\columnwidth]{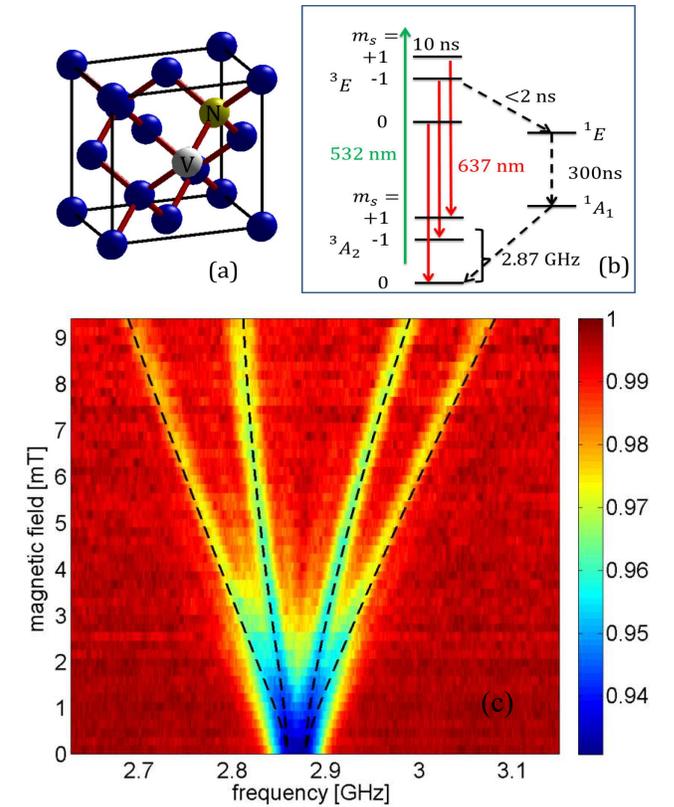}
	 \caption{NV defect. (a) The nitrogen-vacancy defect in diamond. (b) Energy level diagram of the $\text{NV}^{-}$ defect. (c) PL signal (normalized with respect to its maximal value in the plot) showing splitting to four resonances. The dashed lines represent the theoretical prediction based on Eq.~\eqref{eq:nuA} for the case where the magnetic field is applied in the direction $[\cos\theta_{z} \: \sin\theta_{z} \: 0]$, where $\theta_{z}=21\degree$.}
	\label{fig:NVres}
\end{figure}

The nitrogen-vacancy (NV) defect in diamond consists of a substitutional nitrogen atom (N) combined with a neighbor vacancy (V) [see Fig.~\ref{fig:NVres}(a)] \cite{Doherty_1}. Two different forms of this defect have been identified - the neutral $\text{NV}^{0}$ and the negatively-charged $\text{NV}^{-}$. For magnetometry purposes, only the negatively-charged defect is useful, since it provides spin triplet ground and excited states, which can be manipulated using pure optical means [see Fig.~\ref{fig:NVres}(b)].

The technique of diamond magnetometry \cite{Rondin_056503,Maze_644,Balasubramanian_648,Rondin_153118,Acosta_174104,Pham_045021,Balasubramanian_383,Steinert_043705,Vershovskii_1026,Vershovskii_393} is based on optical detection \cite{Gruber_2012,le_121202} of a Zeeman shift of the ground-state spin levels of $\text{NV}^{-}$ defects in a single crystal diamond. The $\text{NV}^{-}$ defects posses relatively long coherence time \cite{Balasubramanian_383} and long energy relaxation time \cite{Harrison_586}. Diamond magnetometry has been employed for studying magnetic resonance imaging \cite{Grinolds_215}, neuroscience \cite{Pham_045021,Hall_1}, cellular biology \cite{Balasubramanian_648,McGuinness_358}, and superconductivity \cite{Bouchard_025017,Waxman_054509}. In addition to magnetometry, $\text{NV}^{-}$ defects in diamond can be used for temperature \cite{Acosta_070801} and strain \cite{Dolde_459} sensing, and for quantum information processing \cite{Maurer_1283,Cai_093030}.

The $\text{NV}^{-}$ defect has $C_{3\text{v}}$ symmetry, leading to a triplet ground and excited states, with optical zero phonon line (ZPL) of
$1.945\text{\,eV}$ (wavelength of $637\text{\,nm})$. The sub-levels $m_\text{S}=0$ and $m_\text{S}=\pm1$ of the ground state triplet $^{3}A_{2}$ are separated by $D=2.87\text{\,GHz}$ in the absence of magnetic field [see Fig.~\ref{fig:NVres}(b)]. Note that $m_\text{S}$ denotes the spin along the $\text{NV}^{-}$ axis [see Fig.~\ref{fig:NVres}(a)]. The excited state $^{3}E$ is a triplet as well, with zero-field splitting of $D_{\text{es}}=1.42\text{\,GHz}$.

The $\text{NV}^{-}$ defect can be excited using green light (laser having wavelength of $532\text{\,nm}$ is employed in the current experiment). Once optically excited in the $^{3}E$ state, the $\text{NV}^{-}$ defect can relax either through the same radiative transition, which gives rise to red photoluminescence (PL), or through a secondary path involving non-radiative intersystem crossing to singlet states, as can be seen in Fig.~\ref{fig:NVres}(b). While optical transitions are spin conserving, these non-radiative crossings are strongly spin selective, as the shelving rate from $m_\text{S}=0$ sublevel is much slower than those from $m_\text{S}=\pm1$. In addition, the $\text{NV}^{-}$ defect decays preferentially from the lowest singlet state towards the ground state $m_\text{S}=0$ sublevel. These spin selective processes allow spin polarization into $m_\text{S}=0$ through optical pumping. Furthermore, since intersystem crossings are non-radiative, the $\text{NV}^{-}$ defect PL is significantly higher when the $m_\text{S}=0$ state is populated. Such a spin-dependent PL response enables the detection of electron spin resonance (ESR) by optical means \cite{Gruber_2012}.

The ground state spin Hamiltonian of the NV$^{-}$ defect in diamond is given by $\mathcal{H}=\mathcal{H}_{\parallel}+\mathcal{H}_{\bot}+hE(S_{x'}^{2}-S_{y'}^{2})$, where the $z'$ direction is taken to be parallel to the $\text{NV}^{-}$ axis, the parallel part $\mathcal{H}_{\parallel}$ is given by $\mathcal{H}_{\parallel}=hDS_{z'}^{2}+g\mu_\text{B}B_{z'}S_{z'}$, the transverse part $\mathcal{H}_{\bot}$ is given by $\mathcal{H}_{\bot}=g\mu_\text{B}(B_{x'}S_{x'}+B_{y'}S_{y'})$, $B_i$ is the magnetic field in the $i$ direction, where $i\in\left\{x',y',z'\right\}$, $S_i$ is the corresponding $3\times 3$ spin $S=1$ matrix, $E\sim5\text{\,MHz}$ and $D$ are axial and off-axial zero-field splitting parameters, respectively, $g\simeq2$ is Land\'{e} g-factor, $h$ is Planck constant and $\mu_\text{B}$ is Bohr magneton \cite{Doherty_1}. By evaluating the eigenvalues of $\mathcal{H}$ using perturbation theory, one finds for the case where $E\ll\gamma_\text{g}B_{z'}\ll D$ that the resonance frequencies $\nu_{\pm}$ corresponding to the transitions $m_\text{S}=0\leftrightarrow m_\text{S}=\pm1$ are given by \cite{Rondin_056503}
\begin{equation}
\nu_{\pm}=D\pm\sqrt{(\gamma_\text{g}B_{\parallel})^{2}+E^{2}}+\frac{3\gamma_\text{g}^2B_\bot^2}{2D},
\label{eq:nuA}
\end{equation}
where $\gamma_\text{g}=g\mu_\text{B}/h\simeq28.024 \, \text{GHz}/\text{T}$ is the electron spin gyromagnetic ratio, $B_{\parallel}=B_{z'}$ and $B_\bot^2=B_{x'}^2+B_{y'}^2$.

The NV defects in a single crystal diamond are oriented along the four lattice vectors $[111]$, $[1\bar{1}\bar{1}]$, $[\bar{1}1\bar{1}]$ and $[\bar{1}\bar{1}1]$. The ODMR data seen in panel (c) of Fig. \ref{fig:NVres} has been obtained with externally applied magnetic field having a vanishing component in the $[001]$ direction, and consequently only 4 resonances are obtained. The dashed lines in Fig. \ref{fig:NVres}(c) represent the frequencies $\nu_{\pm}$ calculated according to Eq.~\eqref{eq:nuA}. In general, the term proportional to $B_\bot^2$ in Eq.~\eqref{eq:nuA} can be disregarded when $B_\bot\ll D/\gamma_\text{g}\simeq0.1 \text{\,T}$. Note, however, that the comparison between the ODMR data seen in Fig. \ref{fig:NVres}(c) and theory yields poor agreement when this term is disregarded. As can be seen from Eq.~\eqref{eq:nuA}, diamond magnetometry becomes insensitive to $B_{\parallel}$ when $B_{\parallel}\ll\gamma_\text{g}^{-1}E\simeq0.2\text{\,mT}$.

\begin{figure}[bt]
	\centering
	\includegraphics[width=0.98\columnwidth]{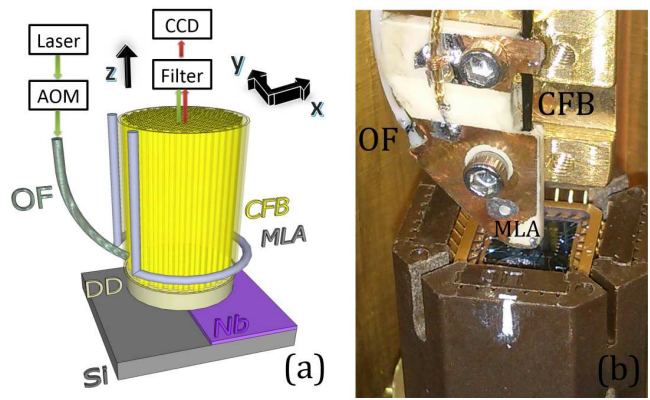}
	\caption{The cryogenic diamond magnetometer. (a) The diamond disk (DD) containing NV defects is glued to the tip of a coherent fiber bundle (CFB) having 30,000 cores, and is brought to contact with the silicon (Si) wafer, which supports the superconducting niobium (Nb) film under study, using a 5-axis positioner. A room temperature optical setup allows optical imaging of the filtered ODMR signal using a charge-coupled device (CCD) camera. An optical fiber (OF) is employed for guiding the laser light into the CFB. Laser intensity is controlled using an acousto optic modulator (AOM). A microwave loop antenna (MLA) is used for applying alternating signals to the NV defects. (b) A photo of the magnetometer probing head and a sample under study.}
	\label{fig:setup}
\end{figure}

Our prototype diamond magnetometer is designed to allow magnetic imaging of an electrically wired sample at cryogenic temperatures. Sketch of the experimental setup is shown in Fig.~\ref{fig:setup}(a). Laser cutting is used to shape a single crystal type Ib diamond into a $15 \,\mathrm{\mu m}$ thick disk having a diameter of $1 \,\text{mm}$. Electron irradiation at $200 \,\text{keV}$ is employed using transmission electron microscope to create defects in the diamond disk. The electron irradiation is followed by annealing at
$900 \,^\circ \text{C}$ for 1 hour and cleaning with boiling Perchloric acid, Nitric acid and Sulfuric acid for 1 hour.

Optical adhesive is then used to glue the diamond disk to the tip of a glass-made coherent fiber bundle (CFB) having 30,000 cores, which allows optical imaging. A magnetometer probing head (see Fig.~\ref{fig:setup}) integrates the CFB, a microwave loop antenna (MLA) and an additional multimode optical fiber (OF), which is used for guiding the laser light at a wavelength of $532 \text{\,nm}$ into the CFB. Note that total internal reflection at the bottom interface of the diamond disk prevents the laser light from reaching the sample under study, avoiding thus undesired heating due to optical absorption by the sample. The electrically wired sample is mounted on a 5-axis piezoelectric positioner having a sub nanometer resolution, allowing reaching contact between the sample and the diamond disk. A charge-coupled device (CCD) camera and a complementary metal-oxide semiconductor (CMOS) one are employed for ODMR imaging of the emitted red photons (long pass dichroic mirror with a cutoff wavelength of $605 \text{\, nm}$ is used).

\begin{figure}
    \includegraphics[width=\columnwidth]{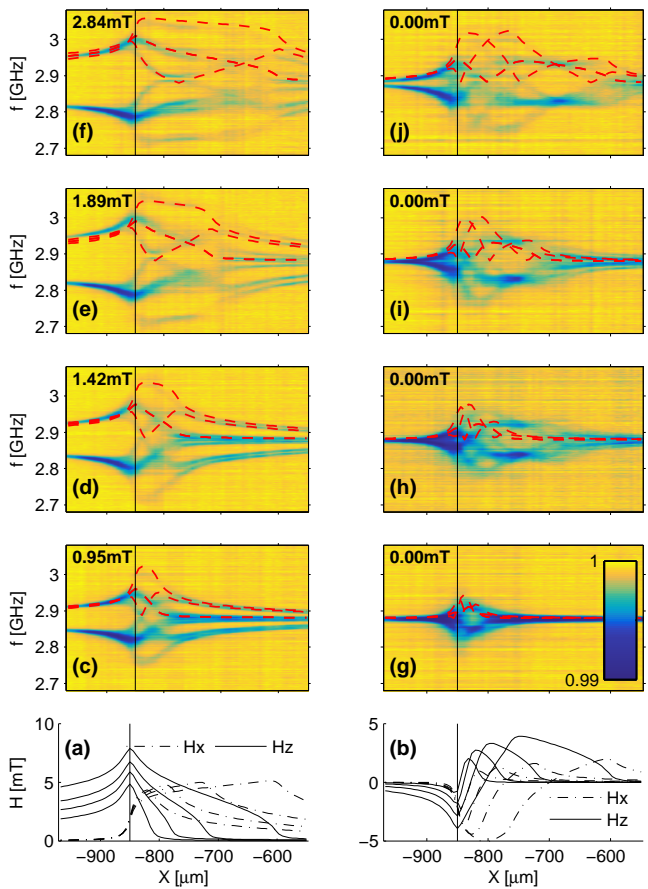}
    \caption{
    ODMR measurements of the magnetic field generated by the shielding currents in the Nb stripe. (c)-(j) ODMR spectrum for various values of the bias field $\Ha$ as a function of distance from the stripe edge, located at $X=-850\,\mathrm{\mu m}$. The two dimensional data were averaged along the $Y$ axis. In (c)-(f) the superconducting sample was first initialized into a virgin state by heating it above $T_\text{c}$ and cooling with no applied field, and then measured with applied bias field of $\Ha$. Then, $\Ha$ was decreased to zero, and the ODMR was measured again [(g)-(j), respectively]. The red dashed lines represent the theoretical prediction that is calculated with the following parameters: distance from the sample surface $Z=7\pm2 \,\mathrm{\mu m}$, $\theta_{z}=45\pm1\degree$, $\theta_{y}=5\pm1\degree$, $B_\text{0}=4\pm0.5~\text{mT}$ and $\Jc=(1.9\pm0.1)\times 10^6\,\text{A/cm}^{2}$\cite{Huebener_344}. The bottom parts of the color-coded plots (which are symmetrical to the top ones) are left without a fit in order to leave raw data clearly visible . Separation to six resonances only (rather than eight) indicates the accuracy of $\theta_{z}$ alignment. The non vanishing distance from the sample surface $Z$ is attributed to the damage depth profile of the electron irradiation \cite{Kim_082410}. The corresponding calculated theoretical magnetic fields are depicted in (a) for (c)-(f) and in (b) for (g)-(j).
    }
    \label{fig:ODMRs}
\end{figure}

A niobium film having a rectangular shape, an area of $5.25\,\text{mm} \times 1.7\,\text{mm}$ and thickness of $d_{\text{Nb}}=500\,\text{nm}$ has been deposited on a high resistivity Si/SiN substrate through a mechanical mask using DC-magnetron sputtering. Magnetometry measurements of the film, whose critical temperature is $9.0\,\text{K}$, are performed at temperature of $4.3\,\text{K}$. A superconducting solenoid is employed for applying a uniform magnetic field perpendicularly to the film. The measured ODMR signal is presented in Fig.~\ref{fig:ODMRs} for various values of the externally applied magnetic field.

The current distribution in a thin film type-II superconductor under applied bias magnetic field $\Ha$ is theoretically evaluated by employing the critical state model \cite{Bean_250,Navau_8201023}. In this model the sheet current density is only allowed to be as high as the critical value $\Jc$. For the case of a constant $\Jc$, the sheet current distribution is found to be given by \cite{Brandt_735,Zeldov_9802}
\begin{equation}
J(x,\Ha,\Jc)=\left\{\begin{array}{lr}
	\frac{2\Jc}{\pi}\arctan\frac{cx}{\sqrt{\xp^2-x^2}} & |x|< \xp \\
    \Jc x/|x| & \xp < |x| < w/2
    \label{eq:J(x)}
\end{array}
\right.,
\end{equation}
where $w=1.7\,\text{mm}$ is the width of the stripe,  $\xp=w/\cosh(\Ha/\Hc)$, $c=\tanh(H_{\text{a}}/H_{\text{c}})$ and $\Hc=\Jc/\pi$ is the critical field. The bias field $\Ha$ is applied along the $z$ axis, and the currents $J(x)$ are along the $y$ axis, as defined in Fig. \ref{fig:setup}.

In general, due to flux trapping the current distribution $J$ is history dependent. For the case where the bias magnetic field is first risen from zero to a maximum value of $\Hu$, and then decreased to $\Hd$, the resulting current distribution is found to be given by \cite{Brandt_735}
\begin{equation}
\Jd(x,\Hd,\Jc)=J(x,\Hu,\Jc)-J(x,\Hd,2\Jc).
\label{eq:Jd}
\end{equation}
After the current distribution is calculated according to Eqs. ~\eqref{eq:J(x)} and ~\eqref{eq:Jd}, the magnetic field in the whole space is computed by integrating the current density with the Biot-Savart kernel.

One of the simplifying assumptions that have been made in the derivation of  Eqs.~\eqref{eq:J(x)} and \eqref{eq:Jd} is that $\Jc$ is independent on the local value of the magnetic induction $B$ \cite{Brandt_735,Zeldov_9802}. More recently, however, various types of dependencies $\Jc(B)$  have been assumed, and the calculation of the current density has been generalized accordingly \cite{Bobyl_184510,DelValle_202506,Davey_1153,Navau_8201023}.  In the so-called exponential model~\cite{Fietz_A335,Young_776} $\Jc(B(x))$ is taken to be given by
\begin{equation}
	\Jc(B(x))=\Jco\exp(-B(x)/\Bo),
	\label{eq:exp}
\end{equation}
where $\Bo$ is a characteristic field and $\Jco$ is the sheet critical current in the low magnetic field limit. To account for the dependence of $\Jc$ on the local value of $B$ we employed the method that has been presented in Ref.~\cite{McDonald_8643} in order to calculate the theoretically predicted magnetic field that is generated by the superconducting film.

The diamond disk was positioned parallel to the sample, in contact with it, and the bias field $\Ha$ was applied in the perpendicular direction. The spatial orientation of the diamond crystal with respect to the cartesian coordinate system that is defined in Fig. \ref{fig:setup} is specified in terms of the unitary transformation $u\left(\theta_{z},\theta_{y}\right)=R_{z}\left(\theta_{z}\right)R_{y}\left(\theta_{y}\right)$, where $R_{y}\left(\theta_{y}\right)$ ($R_{z}\left(\theta_{z}\right)$) represents a rotation around the $y$ ($z$) axis with the rotation angle $\theta_{y}$ ($\theta_{z}$). The transformation is applied to the initial orientation, for which the lattice vectors $[100]$,$[010]$ and $[001]$ are taken to be parallel to the unit vectors $\hat{\textbf{x}}$, $\hat{\textbf{y}}$ and $\hat{\textbf{z}}$, respectively, allowing thus the calculation of the 4 unit vectors pointing in the directions of NV defects in the diamond disk. Next, the frequencies $\nu_{\pm}$ are calculated for each unit vector using Eq.~\eqref{eq:nuA} (see the red dashed lines in Fig. \ref{fig:ODMRs}(c)-(j), which represent the calculated values of $\nu_{+}$).

Comparison between the experimental results and theory is presented in Fig.~\ref{fig:ODMRs}. The ODMR data seen in panels (c)-(f) are obtained after first preparing the sample in the virgin state, and then applying a field $\Ha$, whereas the data seen in panels (g)-(j) are obtained after reducing the field down to zero. The red fit curves are calculated by assuming that $\Jc(B(x))$ is given by Eq.~(\ref{eq:exp}). The only fitting parameters that were not independently measured are $\Jc$ and $\Bo$. In addition to the fitting that is presented in Fig.~\ref{fig:ODMRs}, which is based on the exponential model, other methods have been tested. We found that when the dependence of $\Jc$ on the local value of $B$ is disregarded, i.e. when Eq.~(\ref{eq:J(x)}) is employed for calculating $J(x)$, acceptable agreement with experiment can be obtained in the region of low values of $\Ha$, however the discrepancy becomes significant at high values. Furthermore, the fitting procedure was tested when instead of the exponential model, the so-called Kim's model~\cite{Kim_528} has been employed to determine the dependency $\Jc(B)$. By comparing the results we conclude that the exponential model yields a better (though, not a perfect) agreement with the experimental results.

In summary, the magnetic field generated by shielding currents in a thin superconducting niobium film has been measured. Our diamond magnetometer offers some unique advantages compared with alternative methods that have been previously employed for studying magnetic properties of superconductors~\cite{Bending_11} (see the introductory paragraph above). As was already pointed out above, it allows simultaneous measurement of all three components of the magnetic field vector, and it exploits the effect of total internal reflection to allow low-temperature operation. Furthermore, magnetic field imaging over a large area can be performed without any mechanical scanning~\cite{Hall_1}. The sensitivity of our magnetometer is estimated to be $2\times 10^{-5}\,\text{T}\,\mu\text{m}\,\text{Hz}^{-1/2}$. Further improvements in the design of the magnetometer may enable operation at ultra-low temperatures. Such ability may open the way for a variety of new applications, for example a single-shot quantum state readout of a large array of superconducting Josephson qubits~\cite{Berkley_105014}.

We thank Ran Fischer, Nir Bar-Gill, Eli Zeldov and Rafi Kalish for extremely useful discussions. This work is supported by the Israel Science Foundation and by the  Security  Research Foundation in the Technion.

\end{document}